\documentstyle[12pt]{article}

\begin{document}

\begin{titlepage}

\begin{flushright}
IJS-TP-97/14\\
DTP/97/88\\
NUHEP-TH-97-15\\
CPT-S565.1097\\
\end{flushright}

\vspace{.5cm}

\begin{center}
{\Large \bf Resonant and nonresonant $D^+\to K^-\pi^+l^+\nu_l$ 
semileptonic decays}
\vspace{1.5cm}

{\large \bf B. Bajc $^{a,b}$, S. Fajfer $^{a}$, 
R.J. Oakes $^{c}$ and T.N. Pham $^{d}$}

\vspace{.5cm}

{\it a) J. Stefan Institute, Jamova 39, P.O. Box 3000, 
1001 Ljubljana, Slovenia}

\vspace{.5cm}

{\it b) Department of Physics, University of Durham, 
Durham DH1 3LE, Great Britain}

\vspace{.5cm}

{\it c) Department of Physics and Astronomy, 
Northwestern University, Evanston, Il 60208, 
U.S.A.}

\vspace{0.5cm}

{\it d) Centre de Physique Th\'eorique, 
Centre National de la Recherche Scientifique, UPR A0014, 
Ecole Polytechnique, 91128 Palaiseau Cedex, France}

\end{center}

\centerline{\large \bf ABSTRACT}

\vspace{0.5cm}

We analyse the semileptonic decay $D^+\to K^-\pi^+l^+\nu_l$ 
using an effective Lagrangian developed previously to 
describe the decays $D\to P l \nu_l$ and $D\to V l \nu_l$. 
Light vector mesons are included in the model 
which combines the heavy quark effective Lagrangian and 
chiral perturbation theory approach. The nonresonant and 
resonant contributions are compared. With no 
new parameters the model correctly reproduces the 
measured ratio $\Gamma_{\rm nres}/\Gamma_{\rm nres + res}$. 
We also present useful nonresonant decay distributions. 
Finally, a similar model, but with a modified current 
which satisfies the soft pion theorems at the expense of 
introducing another parameter, is analyzed and the results of 
the models are compared.

\end{titlepage}

\centerline{\bf I. INTRODUCTION}

\vskip 1cm

The experimental result for the nonresonant semileptonic decay mode 
$D^+\to K^-\pi^+l^+\nu_l$ is \cite{frab}

\begin{equation}
\label{e1}
{\Gamma [D^+\to K^-\pi^+\mu^+\nu_\mu ({\rm nonresonant})]\over 
\Gamma [D^+\to K^-\pi^+\mu^+\nu_\mu]} = 0.083\pm 0.029\;.
\end{equation}

\noindent
It is important to understand this result before 
making predictions for the other yet unmeasured $D\to P_1 P_2 l \nu_l$ 
decay modes. Furthermore, this could also be helpful in 
understanding of the $D\to Pl\nu_l$ \cite{anjosone,anjostwo} 
and $D\to Vl\nu_l$ data \cite{anjosthree,kodama}. 

Two main issues arise in the description of $D$ meson 
semileptonic decays. The first potential problem is that the 
$D$ mesons might not be heavy enough for the heavy quark 
effective theory (HQET) \cite{wise} to be very accurate. 
The second possible problem is the 
application of chiral perturbation theory (CHPT) in 
$D\to Pl \nu_l$ and $D\to P_1 P_2 l \nu_l$ decays, where 
the light pseudoscalar mesons can have quite large energies 
\cite{wise,casone,castwo,burdman}. 

In order to investigate the semileptonic $D_{l3}$ decays of $D$ 
mesons we have developed a model \cite{BFO} which accommodates the 
available experimental data using HQET and the CHPT description of 
the heavy and light meson sectors, respectively. 
The experimental data for the semileptonic $D_{l3}$ decays are 
unfortunately not good enough at this time to empirically 
determine the $q^2$ dependence of the form factors. 
What is known experimentally are some branching ratios, 
based on measuring the relevant form factors at some kinematical 
point and {\it assuming} a pole-type behaviour for all the form 
factors. The same assumption 
is also used in many theoretical calculations, for 
example in \cite{castwo} and \cite{stech}. 

In our model \cite{BFO} the {\it vertices} of the processes 
considered are assumed not to change appreciably 
from their value at the zero-recoil point, where they are predicted 
in the heavy quark limit. However, in our model \cite{BFO} 
{\it the complete propagators for the heavy mesons} are used instead 
of the usual HQET propagators. Assuming that these modified Feynman 
rules can be approximately applied to the entire available 
$q^2$ region, one can naturally understand why some form factors 
do have a pole-type behaviour and why others are mainly flat, 
which is also in agreement with 
the predictions of the QCD sum rules analysis \cite{ball}. 
Moreover, in the region where the heavy meson is nearly 
on-shell, where HQET is most reliable, the 
HQET prescription and our model \cite{BFO} almost perfectly overlap, 
providing a simple and consistent picture. 
By calculating the decay widths of all the measured charmed 
meson $D_{l3}$ semileptonic decays we found \cite{BFO} that 
our model, which is a simple modification of the orthodox HQET, 
worked well, providing confidence in extending it to the 
$D_{l4}$ decays.

There are some previous theoretical calculations attempting to describe 
the $D\to K\pi l\nu$ $D_{l4}$ semileptonic decays using the heavy quark 
effective theory (HQET) \cite{burdman,LLWISE,SSS}. Ref. \cite{LLWISE} 
includes only the light pseudoscalars, while for the 
understanding of the experimental data one must also include the 
light vector mesons. The authors of  ref. \cite{SSS} considered 
resonant and nonresonant contributions in the overlapping region 
and indicated that outside the resonant region both contributions 
could be the same order of magnitude. However, no predictions 
were made. In the present investigation we use our simple and 
instructive model \cite{BFO}, which was quite successful in 
describing the $D_{l3}$ decays, to calculate the nonresonant 
contribution to the $D^+\to K^-\pi^+l^+\nu_l$ semileptonic decay. 
The experimental ratio (\ref{e1}) is then found to be 
reproduced without introducing any new parameters, 
which is remarkable, considering the simplicity of the model. 

Since the weak current in the model does not satisfy the 
soft pion limit exactly, we have also investigated a modified 
current which does have the exactly correct soft pion limit for 
comparison, even though the light mesons are not particularly soft 
in the $D_{l4}$ decay. Of course, the strong Lagrangian  automatically 
satisfies the soft pion constraint. 

In Sec. II we briefly summarize the strong Lagrangian describing the 
heavy and light pseudoscalar and vector mesons, based on 
HQET and chiral symmetry. In Sec. III the two weak currents we 
consider are given: the usual one, whose form is based on 
HQET, and another, modified to exactly satisfy the soft pion 
theorems, albeit at the expense of additional parameters. 
In Section IV the expressions describing the $D_{l4}$ semileptonic decays 
are given and used to calculate the resonant and nonresonant decay widths, 
as well as some distributions which might be useful in future 
analyses of experimental data. The results are 
presented in Section V. Finally, a brief summary and a few comments 
are given in Sec. VI.

\vskip 1cm

\centerline{\bf II. THE HQET AND CHPT STRONG LAGRANGIAN}

\vskip 1cm

Our strong interaction Lagrangian \cite{BFO}, which incorporates 
both the heavy meson $SU(2)$ spin symmetry \cite{wisgur}, 
\cite{georgi} and the $SU(3)_L\times SU(3)_R$ chiral 
symmetry, spontaneously broken to the diagonal 
$SU(3)_V$ \cite{bando}, describes the 
heavy and light pseudoscalar and 
vector mesons. A similar Lagrangian, but without the 
light vector octet, was first introduced by Wise 
\cite{wise}, Burdman and Donoghue \cite{burdman}, 
and Yan et al. \cite{yan}. 
It was then generalized to include the light 
vector mesons in \cite{casone}, \cite{kamal}, \cite{ourpaper}.

The light meson sector of the strong Langangian is 

\begin{eqnarray}
\label{defllight}
{\cal L}_{\rm light} = &-&{f^2 \over 2}
\{{\rm tr\,}({\cal A}_\mu {\cal A}^\mu) +
2\, {\rm tr\,}[({\cal V}_\mu - {\hat \rho}_\mu)^2]\}\nonumber\\
& + & {1 \over 2 g_V^2} {\rm tr\,}[F_{\mu \nu}({\hat \rho})
F^{\mu \nu}({\hat \rho})]\;,
\end{eqnarray}

\noindent
where 

\begin{eqnarray}
{\cal A}_\mu &=&{1\over 2}(u^\dagger\partial_\mu u-
u\partial_\mu u^\dagger)\;\;,\;\; 
{\cal V}_\mu ={1\over 2}(u^\dagger\partial_\mu u+ 
u\partial_\mu u^\dagger)\;\;,\\
F_{\mu\nu}(\hat\rho)&=&\partial_\mu\hat\rho_\nu-
\partial_\nu\hat\rho_\mu+[\hat\rho_\mu,\hat\rho_\nu]\;\;,\\
u&=&{\rm exp}(i\Pi /f)\;\;,\;\;
\hat\rho_\mu=i(g_V/\sqrt{2})\rho_\mu\;,
\end{eqnarray}

\noindent
and $\Pi$ and $\rho$ are the usual $3\times 3$ Hermitian pseudoscalar 
and vector matrices. $f=130$MeV is the pseudoscalar decay constant and 
$g_V=5.9$ is determined by the values of the light vector meson masses.

Both the heavy pseudoscalar and the heavy vector 
mesons are described by the $4\times 4$ matrix 

\begin{eqnarray}
\label{defh}
H_a& = & \frac{1}{2} (1 + \!\!\not{\! v}) (P_{a\mu}^{*}
\gamma^{\mu} - P_{a} \gamma_{5})\;,
\end{eqnarray}

\noindent
where $a=1,2,3$ is the $SU(3)_V$ index of the light 
flavours. $P_{a\mu}^*$ and $P_{a}$ annihilate 
spin $1$ and spin $0$ heavy mesons $c \bar{q}_a$ having 
velocity $v$, respectively, and have mass dimension 
$3/2$ so that the Lagrangian 
is explicitly mass independent in the heavy quark 
limit $m_c\to\infty$. Defining 
${\bar H}_{a}=\gamma^{0} H_{a}^{\dag} \gamma^{0} =
(P_{a\mu}^{* \dag} \gamma^{\mu} + P_{a}^{\dag} \gamma_{5})
(1 + \!\!\not{\! v})/2$, we can write the leading order 
strong Lagrangian as 

\begin{eqnarray}
\label{lstrong}
{\cal L}_{\rm even} & = & {\cal L}_{light} +
i {\rm Tr\,} (H_{a} v_{\mu} (\partial^{\mu}+{\cal V}^{\mu}) 
{\bar H}_{a})\nonumber\\
& + &i g {\rm Tr\,} [H_{b} \gamma_{\mu} \gamma_{5}
({\cal A}^{\mu})_{ba} {\bar H}_{a}]
 +  i \tilde\beta {\rm Tr\,} [H_{b} v_{\mu} ({\cal V}^{\mu}
- {\hat \rho}^{\mu})_{ba} {\bar H}_{a}]\nonumber\\
& + &  {\tilde\beta^2 \over 4 f^2 }
{\rm Tr\,} ({\bar H}_b H_a {\bar H}_a H_b)\;,
\end{eqnarray}

\noindent
where ${\cal V}^\mu$ in the heavy meson kinetic term 
makes the derivative covariant and also ensures that the kinetic 
term is chiral invariant, since the heavy meson 
field transforms non-linearly under chiral symmetry 
$SU(3)\times SU(3)$. The Lagrangian (\ref{lstrong}) 
is the most general even-parity Lagrangian to leading order in 
the heavy quark mass and the chiral symmetry limit. 
The first of the four terms in (\ref{lstrong}) is the 
kinetic term for the heavy field $H_a$ and is thus properly 
normalized. The second term represents the strong interactions 
of the pseudoscalar meson field with the heavy meson field. 
The third term gives the interactions of the light vector mesons 
with the heavy field. These terms involve two unknown 
parameters, $g$ and $\tilde\beta$, which are not determined by symmetry 
arguments, and must be determined empirically. 
As we will see below, only the parameter $g$ will be relevant 
in the present investigation. Finally, the last term comes from the requirement 
\cite{casone} that the Lagrangian (\ref{lstrong}) reduces to Wise's 
Lagrangian \cite{wise} in the limit $g_V\to\infty$. The vector field 
$\hat\rho_\mu$ then has no derivatives and can be explicitly integrated out. 
With this requirement the coefficient of the last term is fixed. 
However, this convention is irrelevant for our calculations 
since the vertex with four heavy fields does not 
appear in any diagrams we need to consider. 

We will also need the odd-parity Lagrangian for the 
heavy meson sector. The lowest order contribution 
to this Lagrangian is given by 

\begin{eqnarray}
\label{defoddheavy}
{\cal L}_{\rm odd} & = & i {\lambda} {\rm Tr\,} [H_{a}\sigma_{\mu \nu}
F^{\mu \nu} (\hat \rho)_{ab} {\bar H_{b}}]\;.
\end{eqnarray}

\noindent
The parameter $\lambda$ is a priori free, but we do know that 
it is of the order $1/\Lambda_\chi$ with 
$\Lambda_\chi$ being the chiral perturbation theory 
scale.

\vskip 1cm

\centerline{\bf III. THE WEAK LAGRANGIAN}

\vskip 1cm

The weak Lagrangian for the Cabibbo allowed $D$ meson 
semileptonic decays is given at the quark level by 

\begin{equation}
\label{deflfermisl}
{\cal L}_{\Delta C=\Delta S=1}^{\rm eff}=-{G_F V_{cs}^*\over \sqrt{2}} 
[{\bar l}\gamma_\mu (1-\gamma^5)\nu_l]\;
[{\bar s}\gamma^\mu (1-\gamma^5)c]\;, 
\end{equation}

\noindent
where $G_F=1.17\times 10^{-5}$ GeV$^{-2}$ is the Fermi constant 
and $V_{cs}=0.974$ is the relevant Kobayashi-Maskawa matrix element. 

Of course, as usual, we have to interpret the quark current 
in terms of meson fields. We will present two different models, 
(A) and (B), for the weak part of the Lagrangian:

\vskip 0.5cm

\noindent
Model (A): In the first model, which is based on the HQET approach, 
we assume that the weak current transforms as $({\bar 3}_L,1_R)$ under 
chiral $SU(3)_L\times SU(3)_R$ and is linear in the heavy 
meson field. Using HQET one can then write the most general 
weak current contributing to D meson semileptonic decays 
to leading order in $1/M$ and to the next-to-leading 
order in the chiral expansion as 

\begin{eqnarray}
\label{ja}
j^{(A)}_\lambda = &+& {i\alpha \over 2} J_\lambda u^\dagger \nonumber\\
&-& \alpha_1 J ({\hat \rho}-{\cal V})_\lambda u^\dagger
- \alpha_2 J_\lambda v^\alpha ({\hat \rho}-{\cal V})_\alpha 
u^\dagger \nonumber\\
& + & \alpha_3 J {\cal A}_\lambda u^\dagger
+ \alpha_4 J_\lambda v^\alpha {\cal A}_\alpha u^\dagger \nonumber\\
&+& [J_\lambda v_\alpha - J_\alpha v_\lambda - 
i \epsilon_{\mu\lambda\alpha\beta} J^\mu v^\beta]
[\alpha_1 ({\hat \rho}-{\cal V})^\alpha-\alpha_3 {\cal A}^\alpha] 
u^\dagger\;,
\end{eqnarray}

\noindent
where 

\begin{equation}
J_\lambda={\rm Tr\,}_D[\gamma_\lambda (1-\gamma_5)H]
\end{equation}

\noindent
and

\begin{equation}
J={\rm Tr\,}_D[(1-\gamma_5)H]\;.
\end{equation}

The first term in (\ref{ja}), i.e. the one proportional to 
$\alpha$ ($=f_D\sqrt{m_D}$), is ${\cal O}(E^0)$, while the 
rest is ${\cal O}(E)$ \cite{BFO}. In the process $D\to Pl\nu$ 
\cite{BFO} one takes into account only the 
first term in (\ref{ja}), which is formally the leading term. 
In $D\to Vl\nu$ \cite{BFO} the terms proportional to $\alpha_1$ 
and $\alpha_2$ must be included as well, since the diagrams where 
they appear are of the same order as the diagrams with the terms 
proportional to $\alpha$. Actually, the latter diagram has also a 
$D^*DV$ vertex (\ref{defoddheavy}), which is ${\cal O}(E)$. 

\vskip 0.5cm

\noindent
Model (B): In the decays $D(p_D)\to P_1(p_1)l\nu_l$ or $D(p_D)\to 
P_1(p_1)P_2(p_2)l\nu_l$ one would expect that the part of 
the amplitude proportional to $p_D$ is the most important since 
$p_D^2=m_D^2$ while $p_i^2=m_i^2\ll m_D^2$. The procedure 
described in (A) takes into account the leading $\alpha$ term 
in (\ref{ja}) and neglects the higher terms which contain 
derivatives of the light fields and are thus formally 
next-to-leading order terms. However, what is measured is not 
the amplitude but the matrix elements squared, and because 
the leptons are almost massless the part of the amplitude 
proportional to $(p_D-\sum_ip_i)^\mu$ can not contribute. 
Writing $p_D^\mu=(p_D-\sum_ip_i)^\mu+(\sum_ip_i)^\mu$ one 
sees that the formally large part of the amplitude 
proportional to $p_D^\mu$ contributes to 
the decay width only through the term 
$(\sum_ip_i)^\mu$. So, unless the coefficients $\alpha_{1,2,3}$ in 
(\ref{ja}) are found to be numerically negligible compared 
to $\alpha$ they can contribute comparably, even if formally they 
are next-to-leading order terms. However, the term proportional 
to $\alpha_4$ is of higher order and the terms in the last line of 
(\ref{ja}) do not contribute at all to $D^+\to K^-\pi^+l^+\nu_l$ decay. 
Consequently, the formal procedure described in (A) does not 
take into account these possibly important contributions. 
However, if we assume that the lower dimensional coefficients of higher 
dimensional operators in (\ref{ja}) are naturally smaller (i.e. suppressed 
by powers of some large scale) then we can continue to use the usual 
approach (A), as was done in the past \cite{wise}, \cite{BFO}.

Another potential inadequacy of approach (A) is that the soft pion 
theorems are not satisfied by the weak current. This has its origin 
in the absence of all terms of order $1/m_D$ in (\ref{ja}). 
In our case the soft pion theorem requires that the 
D decay amplitude vanishes in the limit $p_\pi\to 0$ \cite{marshak}. 
The only term in the current (\ref{ja}) that does not satisfy 
this constraint is the first one, i.e. 
$-if_D\sqrt{m_D}v_\lambda Pu^\dagger$, since it does not have a 
derivative acting on the pion field. In the HQET this term comes 
from the term $(f_D/\sqrt{m_D})D(\overleftarrow{\partial}-
{\cal V})_\lambda u^\dagger$, where $D=e^{-im_Dv.x}P$: one 
takes the derivative of only the exponent and neglects the rest, 
which is of higher order in $1/m_D$. However, since 
$-(\partial+{\cal V})_\lambda u^\dagger={\cal A}_\lambda u^\dagger$ 
and the relevant component of ${\cal A}_\lambda$ in our case is 
proportional to the derivative of the pion field, it is clear how to 
modify the weak current in (A) to satisfy the soft pion theorems: 
simply modify (\ref{ja}) by replacing $(P, P^*)$ with $(D,D^*)$ and 
$-im_DDv_\lambda$ with $D(\overleftarrow{\partial}-{\cal V})_\lambda$. 
Keeping only the term relevant for our present purposes explicitly one 
finds this modification of (\ref{ja}) to be

\begin{eqnarray}
\label{jb}
j^{(B)}_\lambda&=&{f_D\over\sqrt{m_D}}D 
(\overleftarrow{\partial}-{\cal V})_\lambda u^\dagger-
2\alpha_3 D{\cal A}_\lambda u^\dagger-
2\alpha_1D(\hat\rho-{\cal V})_\lambda u^\dagger-\nonumber\\
&-&{2\alpha_2\over m_D^2}D(\overleftarrow{\partial}-{\cal V})_\lambda
(\overleftarrow{\partial}-{\cal V})^\alpha(\hat\rho-{\cal V})_\alpha
u^\dagger-f_D^*\sqrt{m_D}D^*_\lambda u^\dagger\;,
\end{eqnarray}

\noindent
which we shall refer to as model (B). Note that this modification of 
the original model (A) based on HQET has come at the expense of the 
appearence of a new parameter, viz. $\alpha_3$. As a result, the 
semileptonic $D\to Pl\nu$ data do not determine the parameter $g$ in 
(\ref{lstrong}) as in model (A) \cite{BFO}, but only a combination of 
$g$ and $\alpha_3$. 

\vskip 1cm

\centerline{\bf IV. THE FORM FACTORS AND DECAY WIDTHS}

\vskip 1cm

Following \cite{LLWISE} we write down the general form 
for the matrix element of the weak current:

\begin{eqnarray}
\label{wwh}
&&<\pi(p_\pi)K(p_K)|\bar s\gamma_\mu(1-\gamma^5)c|D(p_D)>=
ir(p_D-p_K-p_\pi)_\mu\nonumber\\
&&+iw_+(p_K+p_\pi)_\mu
+iw_-(p_K-p_\pi)_\mu-
2h\epsilon_{\mu\alpha\beta\gamma}p_D^\alpha p_K^\beta p_\pi^\gamma\;.
\end{eqnarray}

The form factor $r$ does not contribute to the decay width if 
the lepton mass is neglected and we will not consider it further. 
The following combinations of the remaining three form factors will 
be particularly convenient below:

\begin{eqnarray}
\label{f1}
F_1&=&Xw_++\left[{\beta\over 2}(m_D^2-s_{K\pi}-s_{l\nu}) 
\cos{\theta_K}+\left({m_K^2-m_\pi^2\over s_{K\pi}}
\right)X\right]w_-\;,\\
\label{f2}
F_2&=&\beta (s_{K\pi}s_{l\nu})^{1/2}w_-\;,\\
\label{f3}
F_3&=&\beta X(s_{K\pi}s_{l\nu})^{1/2}h\;.
\end{eqnarray}

\noindent
Here $\theta_K$ is the angle between the kaon three-momentum 
in the $K\pi$ rest frame and the direction of the total momentum 
of the $K\pi$ center of mass in the $D$ rest frame and 

\begin{eqnarray}
s_{K\pi}&=&(p_K+p_\pi)^2\;\;,\;\;s_{l\nu}=(p_D-p_K-p_\pi)^2\;\;,\\
\label{defx}
X&=&{1\over 2}[m_D^4+s_{K\pi}^2+s_{l\nu}^2-2m_D^2s_{K\pi}-
2m_D^2s_{l\nu}-2s_{K\pi}s_{l\nu}]^{1/2}\;,\\
\beta&=&{1\over s_{K\pi}}[s_{K\pi}^2+m_K^4+m_\pi^4-2s_{K\pi}m_K^2-
2s_{K\pi}m_\pi^2-2m_K^2m_\pi^2]^{1/2}\;.
\end{eqnarray}

The $D^+$ meson differential semileptonic decay rate can then be 
written as 

\begin{eqnarray}
\label{dwidth}
{d^3\Gamma\over ds_{K\pi}ds_{l\nu}d\cos{\theta_K}}&=&
{G_F^2|V_{cs}|^2\over (4\pi)^5m_D^3}{X\beta\over 3}\times\nonumber\\
&\times&[|F_1|^2+\sin^2{\theta_K}(|F_2|^2+|F_3|^2)]\;,
\end{eqnarray}

\noindent
where the physical region of phase space is defined 
by $|\cos\theta_K|<1$, $0<s_{l\nu}<(m_D-\sqrt{s_{K\pi}})^2$ 
and $(m_K+m_\pi)^2<s_{K\pi}<m_D^2$. The form factors $F_i$ in 
(\ref{dwidth}) have both resonant and nonresonant parts, which 
we separate by defining

\begin{equation}
\label{division}
F_i=F_i^{r}+F_i^{nr}\;,\;i=1,2,3\;.
\end{equation}

\noindent
The resonant parts are given by 

\begin{eqnarray}
\label{f1res}
F_1^{r}&=&Cg_V\sqrt{m_D\over s_{K\pi}s_{l\nu}}
\left[(m_D^2-s_{K\pi}-s_{l\nu}){\alpha_1\over 2}-
{X^2\over m_D^2}\alpha_2\right]\cos{\theta_K}\;,\\
\label{f2res}
F_2^{r}&=&Cg_V\sqrt{m_D}\alpha_1\;,\\
\label{f3res}
F_3^{r}&=&Cg_V2X\sqrt{m_{Ds*}\over m_D}
{m_{Ds*}f_{Ds*}\over s_{l\nu}-m_{Ds*}^2}\lambda\;,
\end{eqnarray}

\noindent
where

\begin{equation}
\label{c}
C=8\sqrt{\pi s_{l\nu}\over\beta}{\sqrt{m_{K*}\Gamma_{K*}(s_{K\pi})}
\over s_{K\pi}-m_{K*}^2+im_{K*}\Gamma_{K*}(s_{K\pi})}\;,
\end{equation}

\noindent
and

\begin{equation}
\label{kwidth}
\Gamma_{K*}(s_{K\pi})={3\over 2}{g_V^2\over 96\pi}
{\beta^3s_{K\pi}\over m_{K*}}\;.
\end{equation}

\noindent
It is easy to see that for the resonant parts of the form factors, 
in the zero width approximation ($\Gamma_{K*}\to 0$), one obtains 
the previous expressions for the $D^+\to\bar K^{*0}l\nu_l$ decay 
\cite{BFO}.

The nonresonant contributioin to the decay rate is 

\begin{eqnarray}
\label{dwidthnr}
{d^3\Gamma^{\rm nr}\over ds_{K\pi}ds_{l\nu}d\cos{\theta_K}}&=&
{G_F^2|V_{cs}|^2\over (4\pi)^5m_D^3}{X\beta\over 3}\times\nonumber\\
&\times&[|F_1^{nr}|^2+\sin^2{\theta_K}(|F_2^{nr}|^2+
|F_3^{nr}|^2)+\nonumber\\
&&2{\rm Re}(F_1^{nr}F_1^{r*}+\sin^2{\theta_K}
(F_2^{nr}F_2^{r*}+F_3^{nr}F_3^{r*}))]\;.
\end{eqnarray}

The nonresonant form factors will be calculated from the leading 
order Feynman diagrams and given in the next section. Note 
that this nonresonant contribution contains not only the 
nonresonant amplitude itself, but also the interference terms 
with the resonant contribution. 

\vskip 1cm

\centerline{\bf V. RESULTS AND DISCUSSION}

\vskip 1cm

For the calculation of the Feynman diagrams we use the strong 
Lagrangian described in Sec. II and the weak Lagrangian 
from Sec. III for both the currents in models (A) and (B). 
As briefly summarized in the introduction, 
and discussed in detail in 
\cite{BFO}, we use the vertices as given by our Lagrangian, 
assuming that they do not vary appreciably away from the 
maximum recoil point, where the HQET is applicable. However, we use 
the complete heavy meson propagators, instead of the 
HQET approximation. The nonresonant form factors 
can then be straightforwardly calculated from the Feynman diagrams 
with the result:

\begin{eqnarray}
\label{wpnr}
w_{+}^{nr}&=&-{g\over f_K f_\pi}{f_{D*}m_{D*}\sqrt{m_D m_{D*}}\over 
[(p_D-p_\pi)^2-m_{D*}^2]}[1-{p_\pi(p_D-p_\pi)\over m_{D*}^2}]\nonumber\\
&&+{d_+\over 2f_K f_\pi}-{\sqrt{m_D}\over f_Kf_\pi}[\beta{X\over 
m_D^2}\alpha_2 \cos{\theta_K}+{m_K^2-m_\pi^2\over s_{K\pi}}\alpha_1]\;,\\
\label{wmnr}
w_{-}^{nr}&=&{g\over f_K f_\pi}{f_{D*}m_{D*}\sqrt{m_D m_{D*}}\over 
[(p_D-p_\pi)^2-m_{D*}^2]}[1+{p_\pi(p_D-p_\pi)\over m_{D*}^2}]\nonumber\\
&&-{d_-\over 2f_K f_\pi}+{\sqrt{m_D}\alpha_1\over f_Kf_\pi}\;,\\
\label{hnr}
h^{nr}&=&{-2g^2f_{Ds*}m_{Ds*}\sqrt{m_Dm_{Ds*}}\over f_K f_\pi 
[(p_D-p_\pi)^2-m_{D*}^2][(p_D-p_K-p_\pi)^2-m_{Ds*}^2]}\;.
\end{eqnarray}

In \cite{BFO} the parameters $\lambda$, $\alpha_1$ and $\alpha_2$ 
were fitted to correctly reproduce the $D^+\to\bar K^{*0}$ decay. In 
the same reference it was found that the decay mode $D^0\to K^-$ fixes 
the parameters $g$. Due to the nonlinearity of the equations involved 
there are 8 possible sets of these 4 parameters. 
The same values for the decay constants and masses as in 
\cite{BFO} will be used here, namely $f_D=f_{D*}=(0.24\pm 0.05)$ GeV, 
$f_{Ds}=f_{Ds*}=(0.27\pm 0.05)$ GeV \cite{cleo}-\cite{casthree} and 
$m_D=1.87$ GeV, $m_{D*}=2.01$ GeV , $m_{Ds*}=2.11$ GeV \cite{pdg}. 
In deriving (\ref{wpnr})-(\ref{hnr}), and in the following, the approximate 
relation $g_V=m_{K*}/\sqrt{f_Kf_\pi}=5.9$ was used for simplicity 
and will not affect our conclusions. 

The parameters $d_+$ and $d_-$ depend on the choice of the weak current. 
In model (A), $d_+=f_D$ and $d_-=0$, while in model (B) 
$d_+=d_-=f_D-2\sqrt{m_D}\alpha_3$. In the soft pion limit, 
$m_\pi\to 0$ and $p_\pi\to 0$, the constraint

\begin{equation}
\label{wpluswminus}
w_+^{nr}+w_-^{nr}=0
\end{equation}

\noindent
should be satisfied \cite{marshak}. Combining (\ref{wpnr}) and 
(\ref{wmnr}), indeed, one can easily verify that (\ref{wpluswminus}) 
is satisfied in the soft pion limit only in model (B), where $d_+=d_-$, 
but not in model (A), which was the main motivation for including 
model (B), also, in this analysis. 

We note (\ref{wpnr})-(\ref{hnr}) agree with the rsults of Wise, 
et al. \cite{LLWISE}, when the differences off-shell between our 
phenomenological approach of model (A) in which $d_+=f_D$ and $d_-=0$ 
and the exact HQET \cite{LLWISE} are taken into account, 
as discussed above. Next we present numerical results for 
both models (A) and (B).

\vskip 0.5cm

\noindent
In model (A) there are no unknown parameters. As we discussed 
above and showed in \cite{BFO}, there are $8$ possible sets of 
parameters $g$, $\lambda$, $\alpha_1$ and $\alpha_2$ compatible 
with all the $D_{l3}$ decay data. For the $D_{l4}$ decay 
$D^+\to K^-\pi^+l^+\nu_l$ we have calculated the ratio 
$R=\Gamma({\rm nres})/[\Gamma({\rm res})+\Gamma({\rm nres})]$ 
for all eight possible sets and the results are given in Table \ref{tab1}. 
We see from Table \ref{tab1} that all the combinations of 
the allowed values of the input parameters predict a 
ratio $R$ which is consistent with the experimental value (\ref{e1}) 
$R_{\rm exp}=(8.3\pm 2.9)\%$. The errors quoted are due to the 
errors only in the model parameters. Unfortunaletly, due to the large 
experimental error in $R$ the parameters of the model are not restricted 
further. Indeed, the fact that all the $8$ sets of parameters that were 
determined from the $D_{l3}$ data \cite{BFO} give an acceptable 
prediction for $R$ in this $D_{l4}$ decay is remarkable, even given 
the large uncertainties. Perhaps, this merely indicates that $R$ is 
not very model dependent, provided the model fits all the $D_{l3}$ data. 

Since phase space favors smaller $K\pi$ energies the tree 
approximation, even though suspect for large energies, is 
adequate for our purposes. We explicitly checked that the 
nonresonant contribution does not become large at high values 
of $s_{K\pi}$ by calculating the distribution 
$d(\log{\Gamma^{nr}})/d s_{K\pi}$. To illustrate this point 
we show this distribution in Fig. 1, taking as input 
the mean values of the parameters displayed in the $5^{th}$ 
row in Table \ref{tab1}, which predict a value of $R$ near 
to the central value of $R_{\rm exp}$. In Fig. 1 three curves 
are shown: the dashed line is the contribution of the 
nonresonant amplitude squared, which omitts the last line 
in (\ref{dwidthnr}), the dotted line is the contribution 
of the resonance-nonresonance interferece terms, only the 
last line in (\ref{dwidthnr}), while the full line is the 
sum of both these contributions (\ref{dwidthnr}). It is 
clearly seen that the nonresonant contribution to the decay 
width decreases considerably at large $s_{K\pi}$. After 
integrating the curves in Fig. 1 over the invariant mass 
squared of the $K\pi$ system, we found that almost $80\%$ 
of the entire result (\ref{dwidthnr}) comes from the 
nonresonant amplitude, while only approximately $20\%$ 
comes from the resonant-nonresonant interference term. Fig. 1 
is a prediction of our model (A), which can be tested in future 
experiments.

There is another distribution that is certainly interesting 
to compare with experiment to test this model. It is the 
distribution of the charged lepton energy $E_l$ in the $D^+$ 
rest frame. From the kinematics

\begin{equation}
\label{el}
E_l={1\over 2m_D}\left({m_D^2+s_{l\nu}-s_{K\pi}\over 2}+
X\cos\theta_l\right)\;,
\end{equation}

\noindent
where $\theta_l$ is the angle between the charged lepton 
momentum in the lepton pair center of mass system and the 
direction of the lepton pair center of mass momentum in 
the $D^+$ rest frame. $X$ is defined above in (\ref{defx}).

In general, the distribution in $E_l$ is defined by

\begin{equation}
\label{dgdel}
{d\Gamma\over dE_l}=\int ds_{K\pi}ds_{l\nu}d\cos\theta_K 
{d^3\Gamma\over ds_{K\pi}ds_{l\nu}d\cos\theta_K}
{2m_D\over X}
\end{equation}

\noindent 
with \cite{LLWISE}

\begin{eqnarray}
{d^3\Gamma\over ds_{K\pi}ds_{l\nu}d\cos\theta_K}&=&{G_F^2|V_{cs}|^2
\over (4\pi)^5m_D^3}{X\beta\over 2}\times\nonumber\\&\times&\{
{1\over 4}[|F_1|^2+{3\over 2}\sin^2\theta_K(|F_2|^2+|F_3|^2)]-
\nonumber\\&&
{1\over 4}[|F_1|^2-{1\over 2}\sin^2\theta_K(|F_2|^2+|F_3|^2)]
\cos 2\theta_l+\nonumber\\
&&Re(F_2^*F_3)\sin^2\theta_K\cos\theta_l\}\;.
\end{eqnarray}

\noindent
In (\ref{dgdel}) the integration region is defined as in 
(\ref{dwidthnr}), but with the additional constraint 
$|\cos\theta_l|<1$ where $\cos\theta_l$ is given by (\ref{el}). 
The range for $E_l$ is $0<E_l<[m_D^2-(m_K+m_\pi )^2]/(2m_D)$. 

In Fig. 2 this distribution $(d\Gamma^{nr}/dE_l)/\Gamma^{nr}$ 
is given by the solid line. In addition, the contribution of 
the nonresonant amplitude alone is given by the dashed line, 
while the contribution of the interference between the 
resonant and nonresonant amplitude is given by the dotted line.

\vskip 0.5cm

\noindent
Model (B) satisfies the soft pion theorems exactly and is defined 
by the current (\ref{jb}). It has the nonresonant 
amplitudes (\ref{wpnr}-\ref{hnr}), but with the parameters 
$d_+=d_-=f_D-2\sqrt{m_D}\alpha_3$. Only one combination of 
$d_+$ and $g$ is now determined by the $D_{l3}$ semileptonic 
decay data. The relevant $D_{l3}$ form factor \cite{BFO} becomes

\begin{equation}
\label{f1dl3}
F_1(q^2)={1\over f_K}\left(-{d_+\over 2}+gf_{Ds*}{m_{Ds*}
\sqrt{m_Dm_{Ds*}}\over q^2-m_{Ds*}^2}\right)\;.
\end{equation}

\noindent
and the $D_{l3}$ decay width is given by \cite{BFO}

\begin{equation}
\label{dl3}
\Gamma={G_F^2|V_{cs}|^2\over 24\pi^3}\int_0^{(m_D-m_K)^2}dq^2
|F_1(q^2)|^2\left[{(m_D^2+m_K^2-q^2)^2\over 4m_D^2}-m_K^2\right]^{3/2}\;.
\end{equation}

\noindent
After determining the relation between the parameters 
$g$ and $d_+$ from the $D_{l3}$ decay data we then fit 
the remaining parameter from the non-resonant $D_{l4}$ 
decay data. There are, however, $8$ possibilities: 
four possible sets of parameters $\lambda$, $\alpha_1$ 
and $\alpha_2$ for either of the two possible relations 
between $g$ and $d_+$. From (\ref{f1dl3}) and (\ref{dl3}) 
one can see there is a relation of the form 

\begin{equation}
\label{gpm}
g_{\pm}f_{Ds*}={-bd_+\pm\sqrt{b^2d_+^2-
4c(ad_+^2-\Gamma)}\over 2c}\;.
\end{equation}

Labelling the first four sets as given by the 
first, second, third and fourth column in Table \ref{tab1} with 
$g=g_-$ by $1$ to $4$ and the same sets for $\lambda$, 
$\alpha_1$ and $\alpha_2$ but with $g=g_+$ by $5$ to $8$, 
we present their 
predictions as functions of $d_+$ in Fig. 3. All the other 
parameters are the same as in model (A). The experimental band 
between$(R_{\rm exp})_{\rm min}$ and $(R_{\rm exp})_{\rm max}$ around 
the central value (\ref{e1}) is too broad to 
determine the model parameter $d_+=f_D-2\sqrt{m_D}\alpha_3$. 
We can only conclude that model (B) is also compatible with 
the available experimental data (\ref{e1}). Further precision is 
needed to make more conclusive statements and discriminate between 
models (A) and (B).

\vskip 1cm

\centerline{\bf VI. CONCLUSIONS}

\vskip 1cm

We have used the effective model developed in \cite{BFO} to 
calculate the nonresonant contribution to the 
$D^+\to K^-\pi^+l^+\nu_l$ semileptonic decay. 
The result agrees with the experimental data, 
giving further support to the model considered. More precise 
experimental data are, however, needed in order to reduce the 
uncertainties in the parameters and further test the assumptions.

We have also calculated the distributions in the 
$K-\pi$ center-of-mass energy and in the charged 
lepton energy. Both will be useful in comparing with future 
data will provide more sensitive tests of the model. 

Since our original model \cite{BFO} does not obey 
exactly the soft pion theorems, a slightly modified version, 
but involving another parameter, was introduced and explored. 
The predictions of this model are also in agreement with 
the present experimental data. Due to the large experimental 
errors we are unable to distinguish between the two models 
at present.

\vskip 0.5cm

We thank Dave Buchholz for very useful 
discussions. This work was supported 
in part by the Ministry of Science and 
Technology of the Republic of Slovenia (B.B. and S.F.), by the 
British Royal Society (B.B.) and by the U.S. Department
of Energy, Division of High Energy Physics, under grant 
No. DE-FG02-91-ER4086 (R.J.O.).

\newpage

\begin{table}[h]
\begin{center}
\begin{tabular}{|c|c|c|c||c|}\hline
$g$ & $\lambda$ [GeV$^{-1}$] & 
$\alpha_1$ [GeV$^{1/2}$] & $\alpha_2$ [GeV$^{1/2}$] & 
$R$ [$\%$] \\ \hline \hline 
$0.08 \pm 0.09$ & $-0.34 \pm 0.07$ & 
$-0.14 \pm 0.01$ & $-0.83 \pm 0.04$ & $6.4 \pm 0.6$ \\ 
$0.08 \pm 0.09$ & $-0.34 \pm 0.07$ & 
$-0.14 \pm 0.01$ & $-0.10 \pm 0.03$ & $9.0 \pm 1.1$ \\ 
$0.08 \pm 0.09$ & $-0.74 \pm 0.14$ & 
$-0.064 \pm 0.007$ & $-0.60 \pm 0.03$ & $4.7 \pm 0.5$ \\ 
$0.08 \pm 0.09$ & $-0.74 \pm 0.14$ & 
$-0.064 \pm 0.007$ & $+0.18 \pm 0.03$ & $6.4 \pm 0.8$ \\
$-0.90 \pm 0.19$ & $-0.34 \pm 0.07$ & 
$-0.14 \pm 0.01$ & $-0.83 \pm 0.04$ & $8.0 \pm 5.2$ \\ 
$-0.90 \pm 0.19$ & $-0.34 \pm 0.07$ & 
$-0.14 \pm 0.01$ & $-0.10 \pm 0.03$ & $3.0 \pm 3.5$ \\ 
$-0.90 \pm 0.19$ & $-0.74 \pm 0.14$ & 
$-0.064 \pm 0.007$ & $-0.60 \pm 0.03$ & $11.3 \pm 6.2$ \\ 
$-0.90 \pm 0.19$ & $-0.74 \pm 0.14$ & 
$-0.064 \pm 0.007$ & $+0.18 \pm 0.03$ & $5.1 \pm 4.7$\\ \hline 
\end{tabular}
\caption{\label{tab1} The predictions in model (A) for 
the ratio $R=\Gamma^{nr}/(\Gamma^r+\Gamma^{nr})$ in 
the decay $D^+\to K^-\pi^+l^+\nu_l$. Results are given for 
all the values of the input parameters determined from the 
$D_{l3}$ decay data [10]
%\cite{BFO}
. The experimental ratio is 
$R_{\rm exp}=(8.3\pm 2.9)\%$.} \end{center}
\end{table}

\newpage

\noindent
{\bf FIGURE CAPTIONS}

\vskip 0.5cm

\noindent
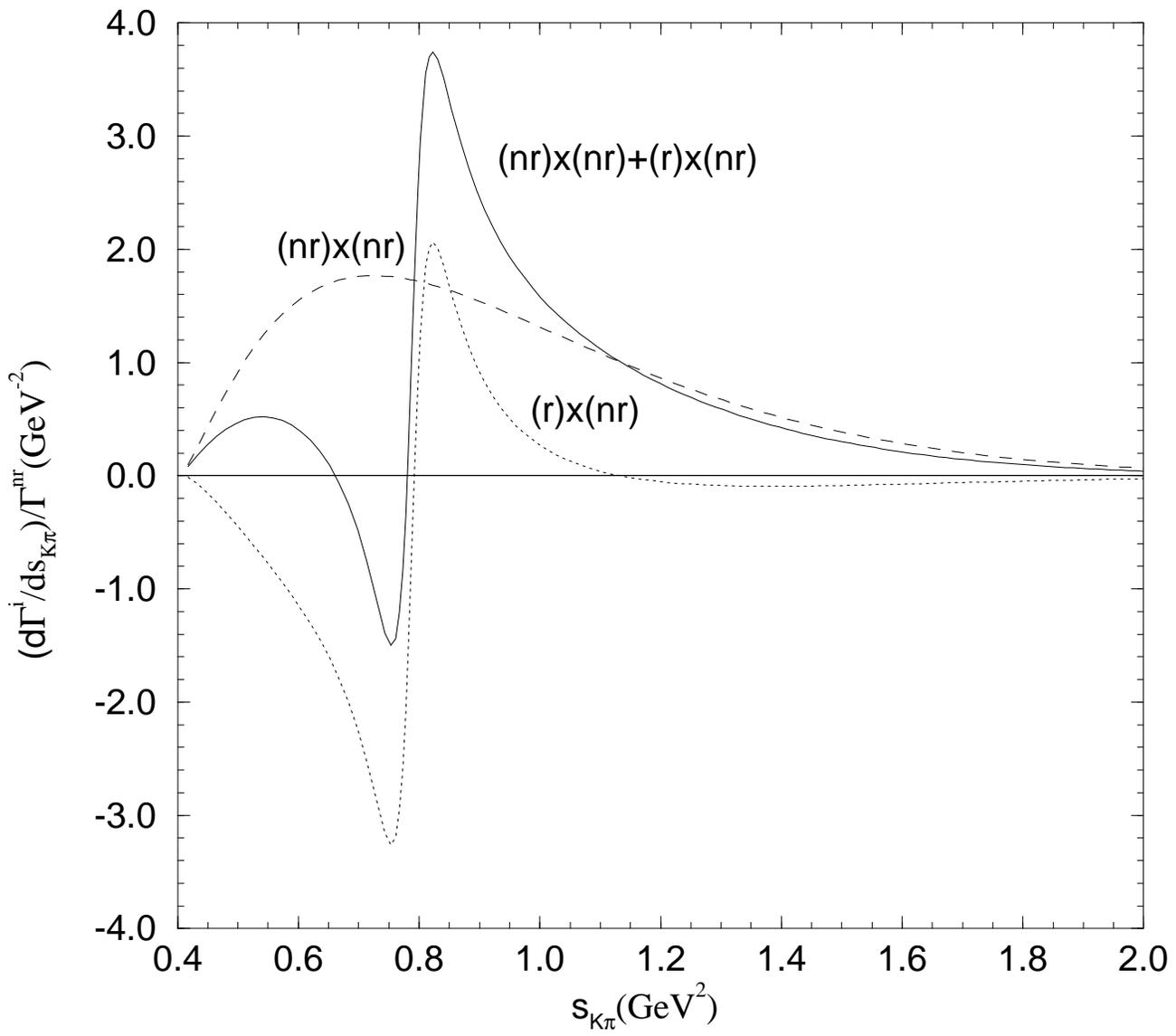
FIG. 1: The contribution to the $s_{K\pi}$ distribution 
for model (A) of the nonresonant and interference terms and their 
sums in Eq. (\ref{dwidthnr}). The dashed line denotes the 
term given by the square of the non-resonant amplitude 
$(nr){\rm x}(nr)$, the dotted line is the interference of the 
resonant and non-resonant amplitudes $(r){\rm x}(nr)$, while 
the full line gives their sum $(nr){\rm x}(nr)+(r){\rm x}(nr)$.

\vskip 0.2cm

\noindent
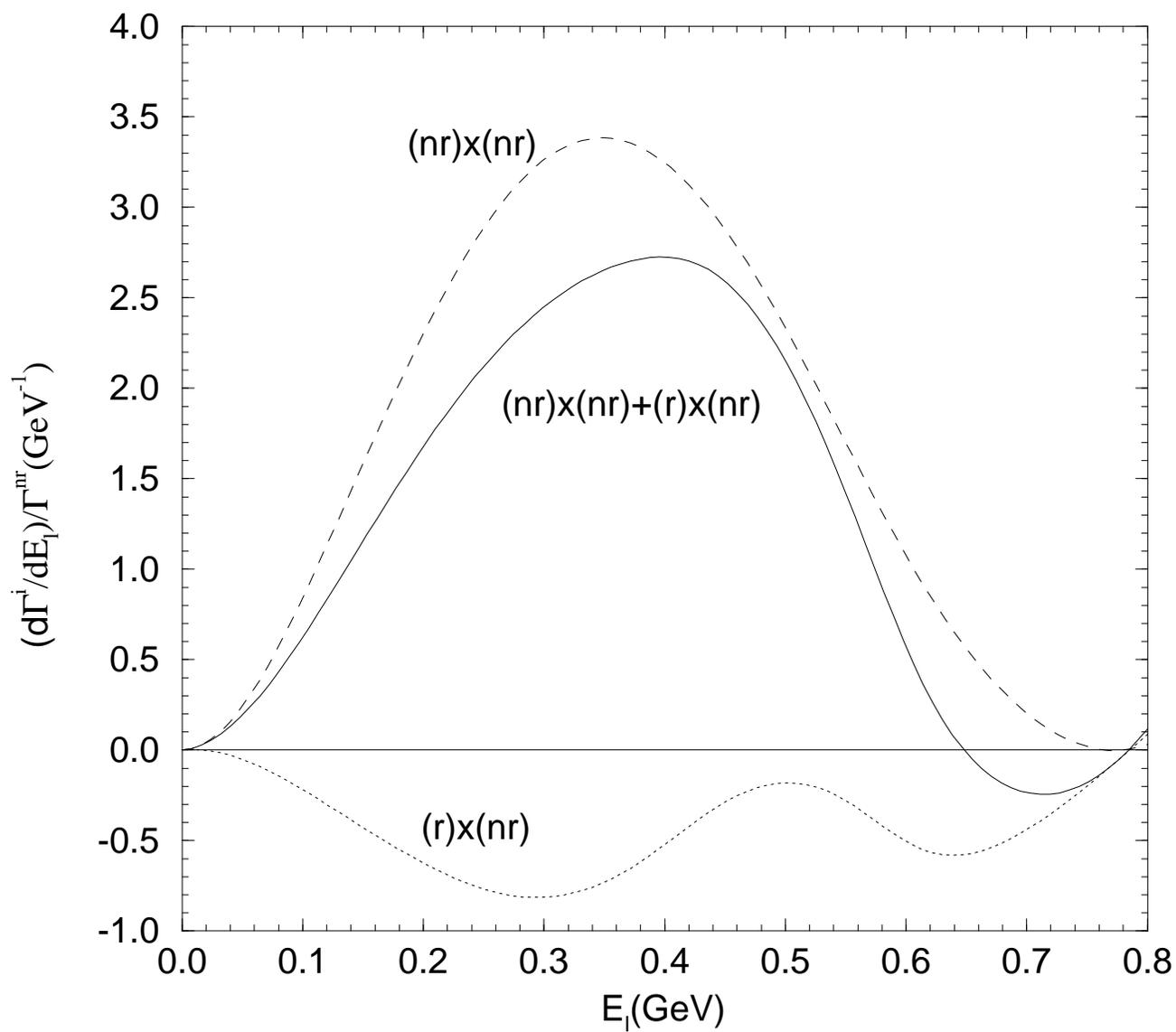
FIG. 2: The charged lepton energy distributions for model (A). 
The dashed line denotes the 
term given by the square of the non-resonant amplitude $(nr){\rm x}(nr)$, 
the dotted line is the interference of the resonant and non-resonant 
amplitudes $(r){\rm x}(nr)$, while the full line gives their sum 
$(nr){\rm x}(nr)+(r){\rm x}(nr)$.

\vskip 0.2cm

\noindent
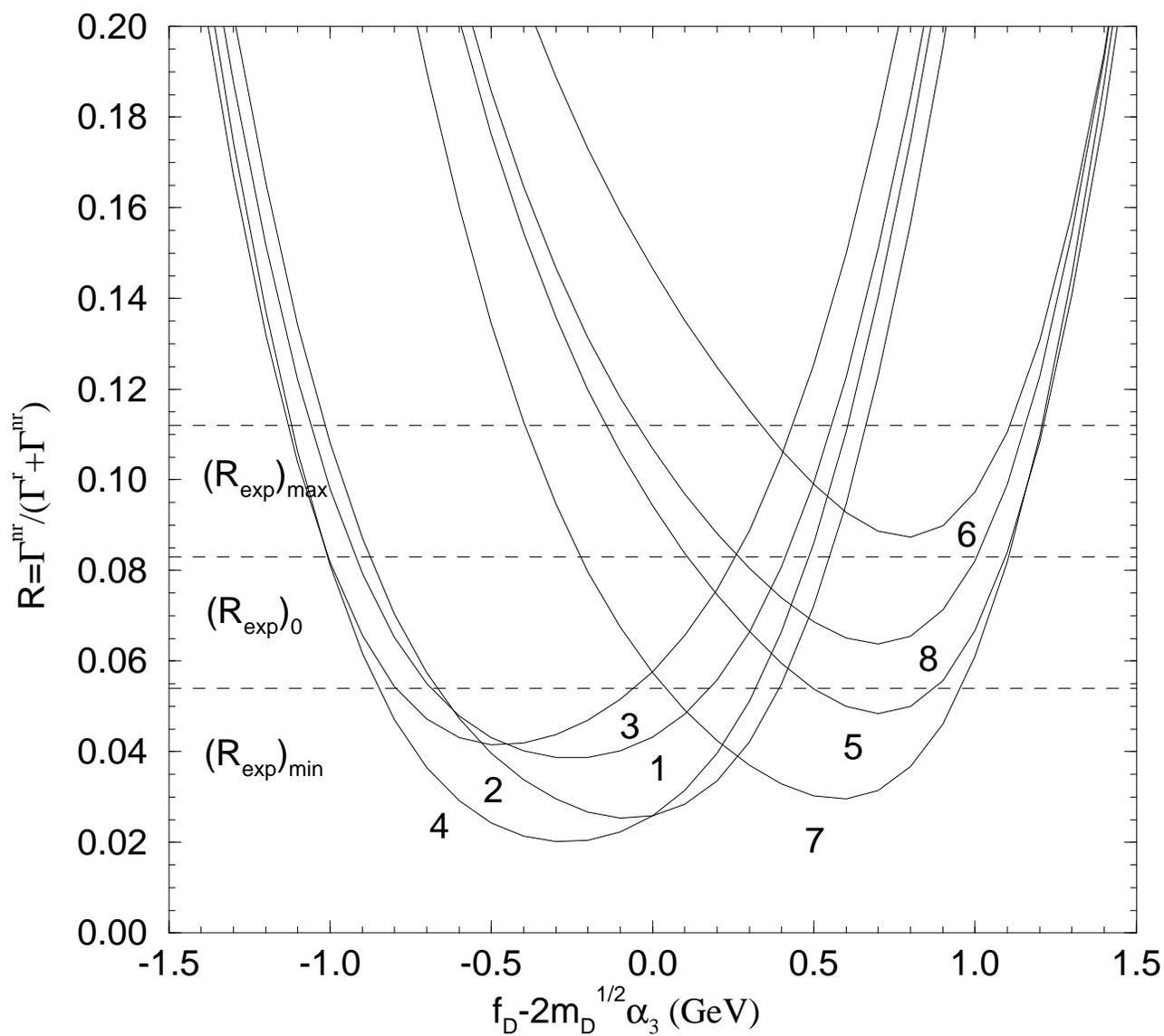
FIG. 3: The eight possible predictions in model (B) for the ratio 
$R=\Gamma^{nr}/(\Gamma^r+\Gamma^{nr})$ as a function of the 
parameter $d_+=d_-=f_D-2\sqrt{m_D}\alpha_3$ 
compared with the experimental central value 
$(R_{\rm exp})_0$ and the values allowed by one standard 
deviation $(R_{\rm exp})_{\rm min}$ and $(R_{\rm exp})_{\rm max}$.

\end{document}